\documentclass[5p]{elsarticle}
\makeatletter
\def\ps@pprintTitle{%
 \let\@oddhead\@empty
 \let\@evenhead\@empty
 \def\@oddfoot{}%
 \let\@evenfoot\@oddfoot}
\makeatother
\usepackage{lineno,hyperref,graphicx}
\usepackage[normalem]{ulem}
\modulolinenumbers[1]
\setpagewiselinenumbers
\journal{Icarus}
\usepackage{multirow}
\usepackage{subfigure}
\usepackage{amsfonts}
\usepackage{amsmath}
\usepackage{booktabs, threeparttable}
\usepackage{array}
\newcolumntype{L}[1]{>{\raggedright\arraybackslash}p{#1}}

\usepackage{color}

\begin{document}

\title{Dissipation and plastic deformation in collisions between metallic nanoparticles}

\author[ucfaddress]{William C. Tucker\corref{mycorrespondingauthor}}
\cortext[mycorrespondingauthor]{Corresponding author.}
\ead{wtucker@ucf.edu}
\author[ucfaddress]{Adrienne R. Dove}
\author[ucfaddress,AMPACaddress]{Patrick K. Schelling}
\address[ucfaddress]{Department of Physics, University of Central Florida, Orlando, FL 32816-2385, USA}
\address[AMPACaddress]{Advanced Materials Processing and Analysis Center, University of Central Florida, Orlando, FL 32804, USA}

\begin{abstract}
Collisions between amorphous Fe nanoparticles were studied using molecular-dynamics simulation. For head-on collisions of nanoparticles with radii $R =$ 1.4 nm, $R =$ 5.2 nm, and $R =$ 11 nm, sticking was observed at all simulated velocities. The results were compared to the description provide{d} by the JKR model. It was found that strong {disagreement} exist{s} between {the predictions of} JKR and the results of the molecular-dynamics simulation due to the presence of additional dissipative processes which strengthen sticking behavior. First, it is demonstrated that very strong dissipation into atomic vibrations occurs during the collision. The dissipation is strong enough to prevent significant rebound of the nanoparticles. Additionally, the morphology of the adhered nanoparticles includes a ``neck'' that increases in radius with increasing collision velocity which results in amplified irreversibility and adhesion. Approximate calculation of the stress during the collision indicates that stress levels are well above typical yield stress values even for low velocity collisions, consistent with the observation of plastic deformation. Furthermore, it is shown that for nanoparticles with $R \leq$ 11 nm, the dominance of surface attraction results in large effective collision velocities and plastic deformation. By obtaining scaling relations for computed quantities, predictions are made for larger nanoparticles up to $R$ $\sim$ 1 $\mu$m. This work provides a new perspective {on} collisional dissipation and adhesion with {an} important connection to {the} modern understanding of tribology and friction. 

\end{abstract}


\maketitle

\section{Introduction}

The collisional dynamics of nano- and micron-scale particles is often described theoretically using the continuum Johnson-Kendall-Roberts (JKR) theory, which describes adhesion between elastic spheres\cite{jkr1971}, with the addition of dissipative processes. Various physical mechanisms have been proposed for dissipation, including elastic waves, viscoelastic dissipation, and dissipative mechanisms related to crack formation\cite{krijt2013,brilliantov2007,wu2003,mesarovic2000,thornton1998,chokshi1993}. While JKR has {typically} been compared favorably with experimental results\cite{blum:2000, blum:1993, poppe:2000}, often the surface energy term is not well known and in some cases is fit to experimental results\cite{kimura:2015,krijt2013}. In addition, no experimental results have clearly demonstrated which dissipative mechanism provides the most suitable description\cite{kimura:2015}. Finally the JKR model is not able to account for plastic deformation, which should become relevant especially at higher collision velocities. 

Recently, atomic-scale simulation has been used to test some assumptions of the JKR model. In Nietiadi et al.\cite{urbassek2017}, molecular dynamics (MD) simulations of collisions between amorphous silica nanoparticles demonstrated strong deviations with the JKR model. Specifically, the critical bouncing velocity was established by simulations to be a factor 3.4 greater than predicted. Furthermore, only sticking was observed for nanoparticles with radii less than 15 nm. This enhanced sticking behavior appeared to be connected to the presence of a larger contact area due to strong plastic deformation, including the generation of filaments, between the nanoparticles. In agreement with these results, it was shown in Quadery et al.\cite{quadery2017} that using MD collisions of silica nanoparticles, the lack of adsorbed OH groups results in strong covalent bonds, with no clear bouncing threshold. However, in Quadery et al.\cite{quadery2017}, only small nanoparticles with radii 2 nm and below were simulated. Nevertheless, existing simulation results suggest strong deviations from JKR model predictions at least for very small nanoparticles. 

The adhesive behavior of powders in turbulent flows is of industrial and theoretical interest as can be seen by the wide range of current research and review articles \cite{millan2012fluidization,MORENOATANASIO2009106,doi:10.1146/annurev-fluid-010814-014644}.  Most of these models utilize the JKR model of contact mechanics in some form.  While Fe nanoparticles might not typically be a material considered for simulating collision dynamics, metallic nanoparticles, and specifically iron nanoparticles, are of interest due to their enhanced catalytic activity, which has been demonstrated for industrial processes \cite{RN23,RN2,RN22,RN5} and in the context of promoting chemical reactions of astrophysical interest\cite{TUCKER2018502}. Iron nanoparticles can even be used to assist in environmental cleanup efforts\cite{Zhang2003}.  
Metallic Fe nanoparticles represent a simple system and are a kind of extreme case where due to dangling bonds at the surface one might expect particularly strong dissipation and adhesion, and thus it represents a sort of ``limiting case'' in collision dynamics which would be interesting to understand.  Velocities were chosen to span a range of regimes but with a special focus on lower velocity collisions from 10 to 500 m s$^{-1}$ which are relevant in the astrophysical context of protoplanetary dust cloud dynamics. For example, in astrophysics, additional dissipative mechanisms and enhanced adhesion could be helpful in surmounting the so-called millimeter bouncing barrier\cite{blum2008}, whereas in a catalytic context the sintering of nanoparticles and consequent loss of surface area leads to a significant drop in effectiveness. Thus, we hope to offer suggestions towards the development of more physically relevant models of adhesion and dissipation for a range of interaction environments.  

In the next section, the basic assumptions of the JKR model applied to nanoparticle collisions with various models of dissipation are briefly described. Next, the atomic-scale simulation methodology used in this paper is presented, followed by a section describing the results of extensive simulations. In the discussion section we attempt to determine scaling relations for relevant physical quantities that can potentially yield predictions for significantly larger length scales, including up to 1 $\mu$m particles. Finally, in the conclusion we tie these results back to potentially relevant applications, including planetary formation, other mineral systems, and previous findings in the field of tribology, including potential relationships to the modern understanding of Amonton's laws of friction.

\section{JKR theory and treatment of dissipation}

In the JKR theory\cite{jkr1971}, the energy associated with adhesion between two identical spheres of radius $R$ with reduced radius $R^*=\frac{R}{2}$ depends on the interfacial energy and the elastic strain energy. The interfacial energy $U_S$ depends on the surface energy $\gamma$ and contact radius $a$,

\begin{equation} \label{jkrsurf}
U_S = -2 \pi a^2 \gamma,
\end{equation}
while {the} elastic strain energy $U_E$ is given by

\begin{equation} \label{jkrelastic}
	U_E=\frac{E^* a^3}{3 R^*} \Big[ \delta \Big( \frac{3 \delta R^*}{a^2} -1 \Big) -\frac{a^2}{5 R^*} \Big( \frac{5 \delta R^*}{a^2} -3 \Big) \Big]					.
\end{equation}
For two identical spheres, the {combined} elastic modulus $E^*$ is defined by

\begin{equation} \label{Estar}
	E^*=\frac{E}{2 ( 1 - \nu )^2}	,	
\end{equation}
where $E$ is the Young's modulus and $\nu$ is the Poisson ratio {of the material in bulk}. The quantity $\delta$ is the length associated with compression of the two sphere{s} in contact. Specifically, for two identical spherical objects with radius $R$ whose center-of-mass coordinates are separated by a distance $d$, the compression is given by

\begin{equation} \label{deltatwor} 
\delta=2 R - d.
\end{equation}
For a particular compression $\delta$, JKR theory predicts that the system will optimize the contact radius $a$ to minimize the total energy $U_{JKR} = U_S + U_E$, resulting in a relationship between the compression $\delta$ and contact radius $a$,

\begin{equation} \label{deltajkr}
	\delta=\frac{a^2}{R^*}-2\sqrt{\frac{\pi \gamma a}{E^*}}.
\end{equation}
With this assumption, the JKR theory results in a force

\begin{equation} \label{Fjkr}
	F_{JKR} = \frac{4 E^* a^3}{3 R^*} - 4 \sqrt{\pi \gamma E^* a^3}.
\end{equation}
The point where $F_{JKR}$ vanishes yields the equilibrium contact radius and length of compression,

\begin{equation} \label{aeq}
	a_{eq} = \Big( \frac{9 \pi \gamma R^{*2}}{E^*} \Big)^{ 1 / 3 },
\end{equation}
\begin{equation} \label{deltaeq}
	\delta_{eq} = \Big( \frac{3 \pi^2 \gamma^2 R^*}{E^{*2}} \Big)^{ 1 / 3 }.
\end{equation}

As can be seen in the above expressions, the JKR theory applied to collisions predicts that equilibrium is attained with a contact area and compression {(and thus an adhesive energy)} that are independent of the collision velocity. These assumptions are not valid if significant plastic deformation occurs.

In order for adhesion to occur, complete dissipation of the incident kinetic energy is required. Dissipation always involves generation of internal thermal energy, while for large enough collision velocities, plastic deformation, generation of coordination defects, and melting can occur. Several attempts have been made to add dissipation to existing JKR-based adhesion models. For example, in Krijt at al.\cite{krijt2013}, models of viscoelastic growth of cracks, bulk viscoelastic dissipation, and plastic deformation were used to describe dissipation. The resulting model provides theoretical predictions for the coefficient of restitution and the sticking velocity. In Chokshi et al.\cite{chokshi1993}, dissipation into elastic waves was used to describe dissipation, with the critical sticking velocity predicted from the requirement that the energy dissipation in elastic deformations be greater than the energy required to separate the nanoparticles.  However, while many models for dissipation exist, no results have been reported which validate any particular dissipation model.

\section{Molecular-Dynamics simulation approach}

In the present article, the detailed atomic-scale mechanisms for dissipation are described using molecular dynamics (MD) simulation. This approach includes all length scales for dissipation, including vibrational modes with wavelengths comparable to the separation between the atoms. The advantage of this approach is that there is no requirement for a physical model with assumptions, but rather all atomic degrees of freedom are explicitly described.

\begin{figure}
\caption{Top right: Atomic structure of the large amorphous Fe nanoparticle with approximate radius $R$ = 11 nm and $N$ = 470561 atoms.  Bottom left: Radial distribution function of the depicted nanoparticle.}
\label{fig:nprdf}
\centering
\includegraphics[width=\linewidth]{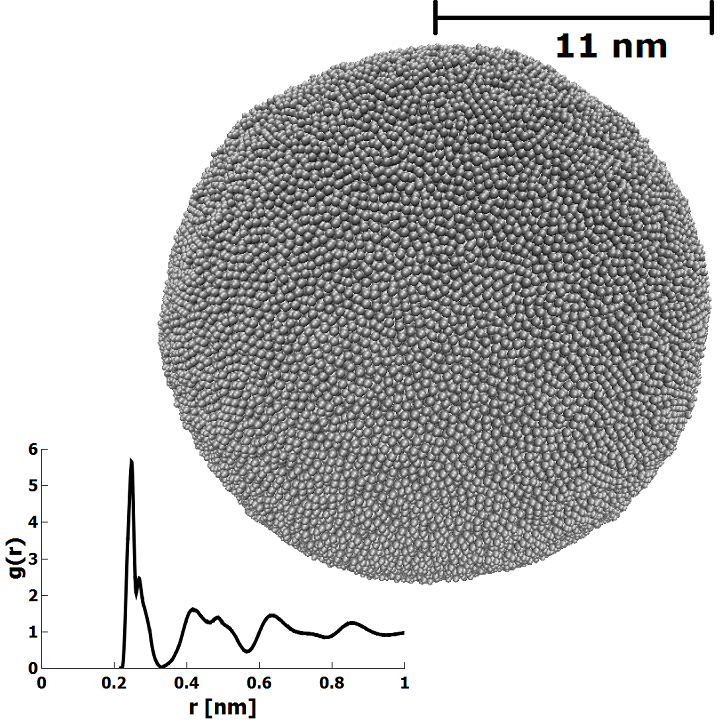}
\end{figure}

The LAMMPS simulation code\cite{plimpton1995} with the embedded-atom method (EAM) potential for Fe from Mendelev et al. (potential 2)\cite{mendelev2003} was used. Visual renders of atomic structures were produced using the Visual Molecular Dynamics (VMD) software package\cite{hump96}. The simulations were performed with an MD time step of 0.25 fs, small enough to ensure energy conservation, and much smaller than the timescale of the highest interatomic vibrational frequency in the system. Three sizes of amorphous nanoparticles were generated: a small nanoparticle with N = 1024 atoms and radius R = 1.4 nm, a medium nanoparticle with N = 50286 atoms and radius R = 5.2 nm, and a large nanoparticle with N = 470561 atoms and radius R = 11 nm. The Fe nanoparticles were melted in a constant temperature ensemble by increasing the temperature to T = 2200 K. This was followed by a slow anneal to T = 5 K.  Melting was performed over 100 ps for the small nanoparticle, 120 ps for the medium nanoparticle, and 240 ps for the large nanoparticle.  Annealing times were identical to melting times.  In the top right of Figure \ref{fig:nprdf}, the atomic structure of the large amorphous Fe nanoparticle is depicted. After the annealing process was complete, the Fe-Fe radial distribution function (RDF) was calculated to ensure that the nanoparticles were amorphous.  The RDF of the large nanoparticle is depicted in the bottom left of Figure \ref{fig:nprdf}. From the RDF data, the coordination number was determined by integration to the first minimum of the RDF at 0.33 nm. The calculated Fe-Fe coordination number of 13.2 is somewhat higher than for an FCC crystal due to the fact that the first minimum in the RDF is significantly greater than the Fe-Fe bond lengths $\sim$ 0.25 nm in either BCC or FCC iron.

The radii of the nanoparticles were determined first by computation of the $T = 0$ K density of bulk amorphous Fe, and then assuming a spherical shape for the nanoparticles. For a nanoparticle with mass $m$ and mass density $\rho$, the radius is defined by

\begin{equation} \label{rdensity}
	R = \Big( \frac{3 m}{4 \pi \rho} \Big)^{ 1 / 3 }.	
\end{equation}
Using this expression, with the computed mass density $\rho =$ 7.82 g cm$^{-3}$ for amorphous Fe, the small nanoparticle with $N =$ 1024 corresponds to a radius $R =$ 1.4 nm. For the medium nanoparticle with $N =$ 50286, the radius is $R =$ 5.2 nm. Finally, for the large nanoparticle with $N =$ 470561 we obtained the radius $R =$ 11 nm.  Given the radii of the nanoparticles, the surface energy $\gamma$ was determined using the computed energy of the nanoparticles and that of the bulk amorphous Fe solid. In each case, the surface energy was computed to be nearly $\gamma =$ 0.09 eV $\text{\AA}^{-2}$, indicating that there was no size-dependence to the surface energy.

Collisions between same-sized nanoparticles with relative velocities between 10 m s$^{-1}$ and 3000 m s$^{-1}$ were simulated.  These velocities were chosen to span the entire range from moderately slow collisions up through collisions with enough kinetic energy to ensure a liquid final state. We cast a special focus on lower velocity collisions up to 500 m s$^{-1}$, as these velocities were found to result in minimal large-scale plastic deformation of the nanoparticle structure away from the inter-particle contact.  For each relative velocity, multiple collisions were simulated for different random rotations of the two nanoparticles. For the small and medium nanoparticles, 30 simulations were performed at each velocity. For the large nanoparticle, 3 simulations were performed at each velocity. After relaxation, both nanoparticles were given a desired translational velocity which resulted in a head-on collision. Simulations were continued for at least 50 ps after the collision until equilibrium was achieved. The center-of-mass coordinates of the two nanoparticles were monitored to determine whether the nanoparticles adhered together. For each simulated nanoparticle size and incident velocity only sticking was observed, with no incidents of bouncing behavior. 

To compute the work of adhesion $W_{adh}$ from the results of the irreversible collision simulations, the approach first described in Quadery et al.\cite{quadery2017} was used. The basic assumption is that nanoparticle collisions result in a change in potential energy and thermal excitation.  If the system remains in a solid state, it is reasonable to assume that the thermal energy can be described using the equipartition theorem applied to a system of harmonic oscillators.  The work of adhesion $W_{adh}$ was therefore computed according to, 

\begin{equation} \label{wadh}
	W_{adh} = 3 N_{tot} k_B (T_f-T_i)-K_{trans},
\end{equation}
where $N_{tot}=2N$ is the total number of atoms in the simulation, $T_f$ and $T_i$ are the computed temperatures before and after the collision, and $K_{trans}$ is the translational kinetic energy before the collision. In determining $T_f$ and $T_i$, the kinetic energy in the center-of-mass reference frame of each nanoparticle was used.  The physical meaning of $W_{adh}$ is that it approximately corresponds to the work required at $T = 0$ K to adiabatically separate the nanoparticles by breaking bonds at the interface. While the interpretation of $W_{adh}$ is clear when adhesion occurs with minimal disruption of the surfaces, in instances with large plastic deformation or melting of the nanoparticles, significant amounts of energy can be stored in disruption of the atomic structure, and the concept of adiabatically separating the nanoparticles after the collision is somewhat ill-defined. Nevertheless, $W_{adh}$ is calculated using Eq. \ref{wadh} for all collisions including those where the physical interpretation is more complicated.

\section{Results}

Each simulated head-on collision resulted in sticking. Fragmentation was not observed at any of the simulated velocites or radii.  For very large velocities, full melting was observed: for v$_{rel}$ = 2750 m s$^{-1}$, T$_f$ was around 1900K, and for v$_{rel}$ = 3000 m s$^{-1}$, T$_f$ was around 2200K, comparable to the solid to fully liquid transition temperature T$_{liq}$ $\sim$ 1900 K of a single $R$ = 1.4 nm nanoparticle melted in an ancillary MD simulation. Consequently, there is no velocity range where bouncing of solid nanoparticles occurs, although at very large velocities either vaporization or splashing will occur. It may be that bouncing events are not impossible, but that they are at least extremely rare. This is in significant contrast to previous results for SiO$_2$ nanoparticles in Quadery et al.\cite{quadery2017} where significant instances of bouncing were observed. Bouncing might also occur for very large nanoparticle sizes or when collisions are not exactly head-on. 

\begin{figure}
\caption{Computed values of $W_{adh}$ for nanoparticles with $R$ = 1.4 nm plotted as a function of relative collision velocity. For each simulated collision velocity, a visualization of a typical structure is also included.  {Lines indicate predictions for JKR elastic energy U$_E$ (negative of Eqn. \ref{jkrelastic}, short dashed line), surface energy U$_S$ (negative of Eqn. \ref{jkrsurf}, long dashed line), and total energy U$_{JKR}$ (sum of the negative of surface and elastic contributions, solid line).}  The inset plots values of $W_{adh}$ for higher velocities.  Units of the inset are identical to units of the main plot, and the inset does not include JKR predictions to reduce visual clutter.}
\label{fig:wadhsmall}
\centering
\includegraphics[width=\linewidth]{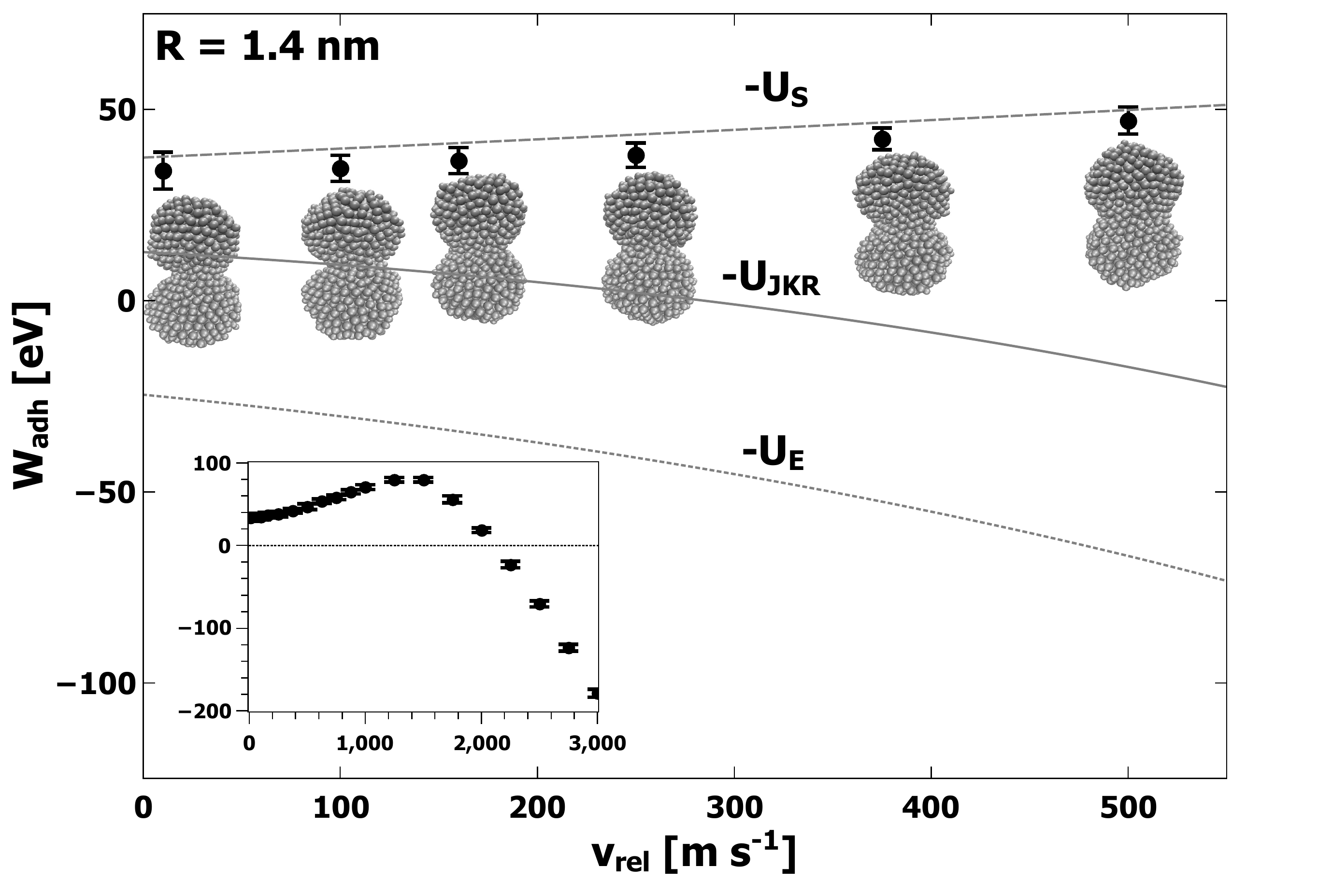}
\end{figure}

\begin{figure}
\caption{Computed values of $W_{adh}$ for nanoparticles with $R$ = 5.2 nm plotted as a function of relative collision velocity. For each simulated collision velocity, a visualization of a typical structure is also included.  The lines and inset are described in the caption of Figure \ref{fig:wadhsmall}.}
\label{fig:wadhmed}
\centering
\includegraphics[width=\linewidth]{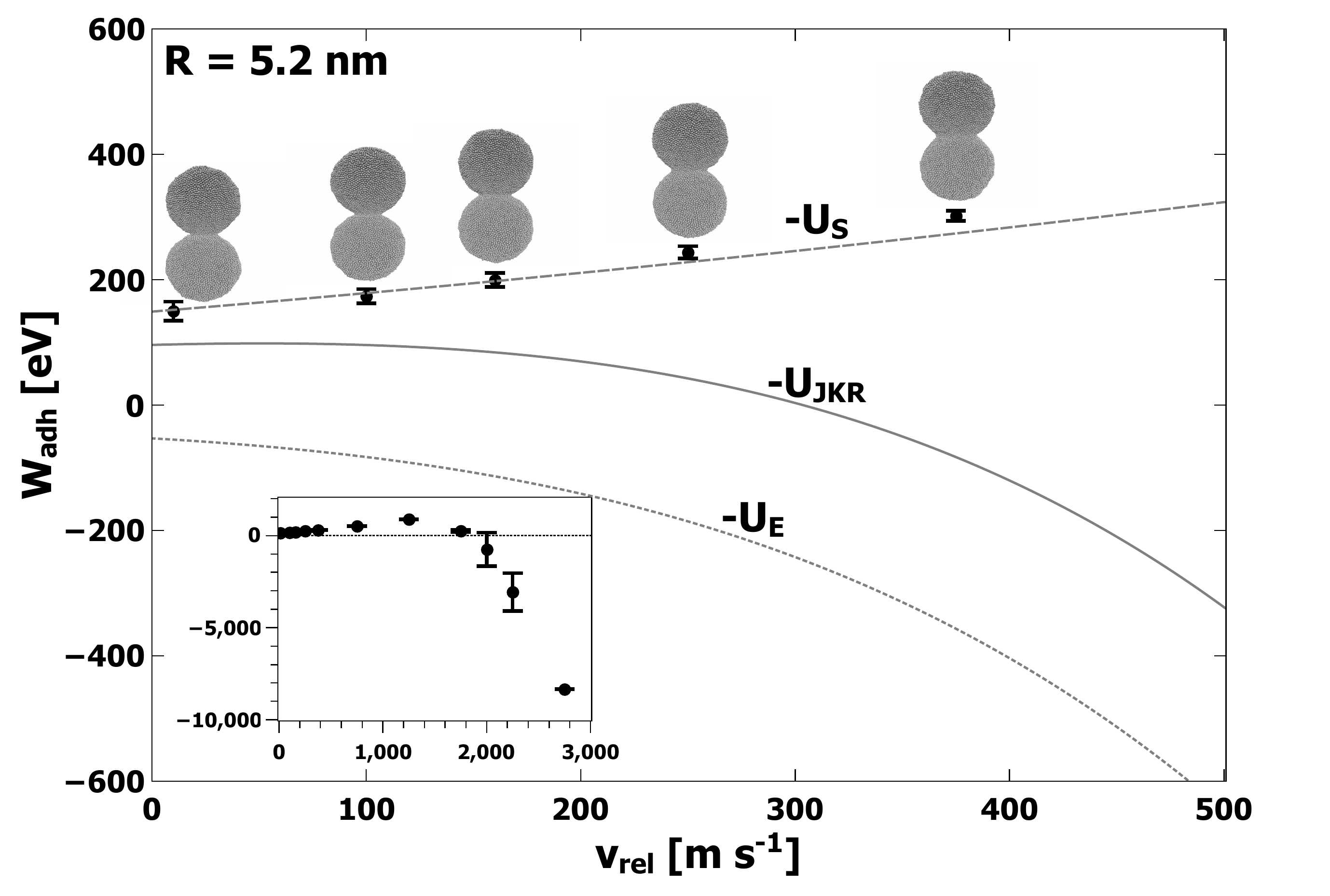}
\end{figure}

\begin{figure}
\caption{Computed values of $W_{adh}$ for nanoparticles with $R$ = 11 nm plotted as a function of relative collision velocity. For each simulated collision velocity, a visualization of a typical structure is also included.  The lines are described in the caption of Figure \ref{fig:wadhsmall}.  Simulations for this size nanoparticle did not extend into higher velocities due to the extensive computational resources required.}
\label{fig:wadhbig}
\centering
\includegraphics[width=\linewidth]{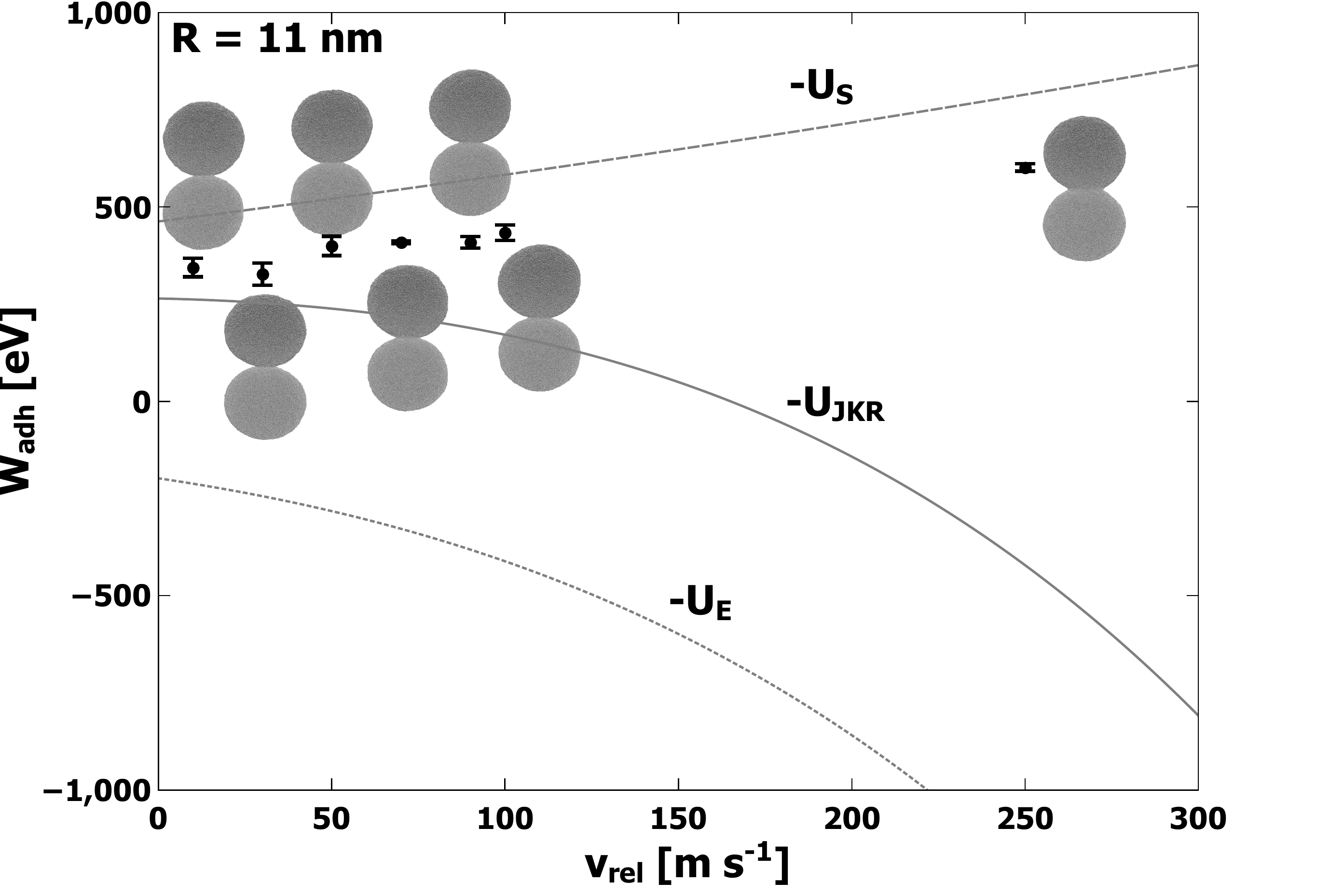}
\end{figure}

In Figures \ref{fig:wadhsmall}, \ref{fig:wadhmed}, and \ref{fig:wadhbig}, we show the computed average values of $W_{adh}$ as a function of collision velocity $v_{rel}$ for lower velocities, with an inset showing the data for all velocities. The error bars represent the standard deviations obtained from all simulations performed at each value of $v_{rel}$. Along with the computed values, an image of a typical atomic structure for the adhered nanoparticles is shown for each collision velocity. Each structure depicted represents the final structure attained after equilibration. Predictions made by JKR for elastic, surface, and total energies are denoted by lines on Figures \ref{fig:wadhsmall}, \ref{fig:wadhmed}, and \ref{fig:wadhbig},. Comparison to JKR theory will be presented in the next section.

From the data {plotted} in Figures \ref{fig:wadhsmall}, \ref{fig:wadhmed}, and \ref{fig:wadhbig}, several consistent trends emerge which are independent of either the model or the size of the nanoparticle. First, the computed value of $W_{adh}$ increases gradually with increasing collision velocity for values of $v_{rel}$ up to at least 1000 m s$^{-1}$. Corresponding to these cases, the structures in Figures \ref{fig:wadhsmall}, \ref{fig:wadhmed}, and \ref{fig:wadhbig}, show an apparent contact area which increases with $v_{rel}$. For larger values of $v_{rel}$ (i.e. significantly past 1000 m s$^{-1}$), the computed $W_{adh}$ begins to decrease and eventually becomes negative. In this regime, the atomic structure appears less like two adhered nanoparticles and progressively more like a single deformed nanoparticle, generally elliptical in shape, eventually becoming spherical at the highest simulated values of $v_{rel}$. Based on calculations of the self-diffusion coefficient, values of $v_{rel}$ above about 2000 m s$^{-1}$ were in a liquid state when fused into a spherical shape. For the largest nanoparticles with radius $R =$ 11 nm, only velocities $v_{rel} =$ 250 m s$^{-1}$ and below were simulated, and the increase in $W_{adh}$ with $v_{rel}$ is apparent but less dramatic. The simulations for the large nanoparticles with $R =$ 11 nm did not extend into the regime where large deformation and melting occurs due to the extensive computational resources required.

The general trends can be understood in a simple way. The changing nature of the contact area is indicative of significant plastic deformation. Specifically, at lower velocities (i.e. below $\sim$ 1000 m s$^{-1}$) the gradual increase in $W_{adh}$ with increasing $v_{rel}$ is due to plastic deformation which allows for greater contact between the two surfaces. Although this could occur by elastic deformation, it is important to note that elastic deformation would also involve significant elastic strain energy (see analysis in next section for further discussion). At higher velocities (above $\sim$ 1000 m s$^{-1}$), the decrease in $W_{adh}$ is due to the complete fusing of the two nanoparticles along with the generation of significant numbers of coordination defects, and possibly strain energy. The coordination defects correspond to stored energy, and thus result in negative contributions to $W_{adh}$. In these instances, separation of the two nanoparticles would result in nanoparticles with markedly different structures than before the collision. Nevertheless, $W_{adh}$ is a measure of the internal potential energy of the system with respect to the isolated nanoparticles before the collision. Finally, at the highest simulated values of $v_{rel}$, the change in internal energy is large enough to result in a phase transition to the liquid state along with very large negative values of $W_{adh}$.

\begin{figure}
\caption{Computed values of $\eta$ for relative collision velocities up to 500 m s$^{-1}$ for all three sizes of nanoparticles plotted as a function of relative collision velocity.  Blue diamonds represent data for R = 1.4 nm, red squares represent data for R = 5.2 nm, and black circles represent data for R = 11 nm.}
\label{fig:etavel}
\centering
\includegraphics[width=\linewidth]{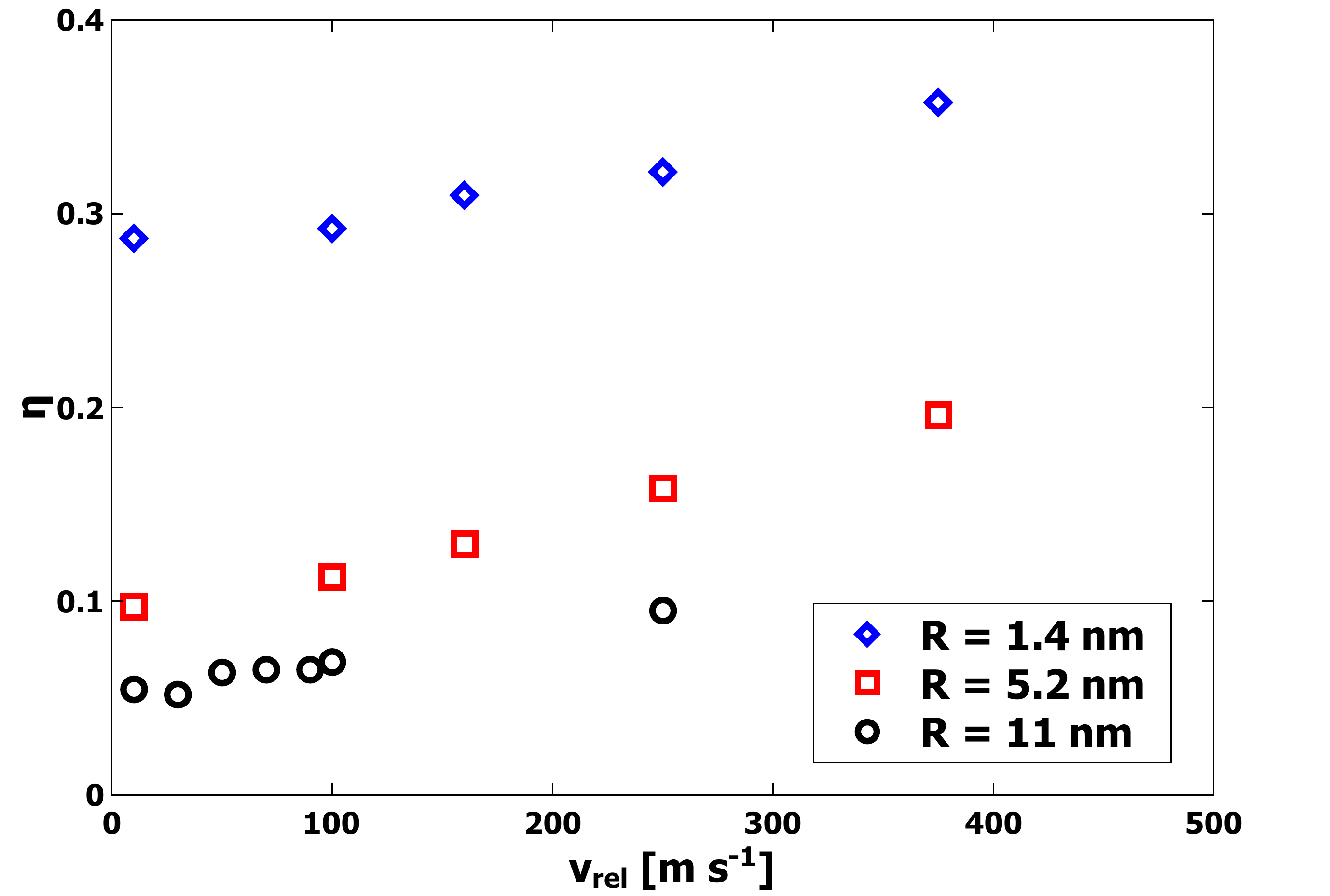}
\end{figure}

\begin{figure}
\caption{Computed values of the compression length $\delta$ for relative collision velocities up to 500 m s$^{-1}$ for all three sizes of nanoparticles plotted as a function of relative collision velocity. Blue diamonds represent data for R = 1.4 nm, red squares represent data for R = 5.2 nm, and black circles represent data for R = 11 nm.  JKR theory  predicts values of $\delta_{eq}$ = 0.23 nm, 0.35 nm, and 0.43 nm for the small, medium, and large nanoparticles, respectively.}
\label{fig:deltavel}
\centering
\includegraphics[width=\linewidth]{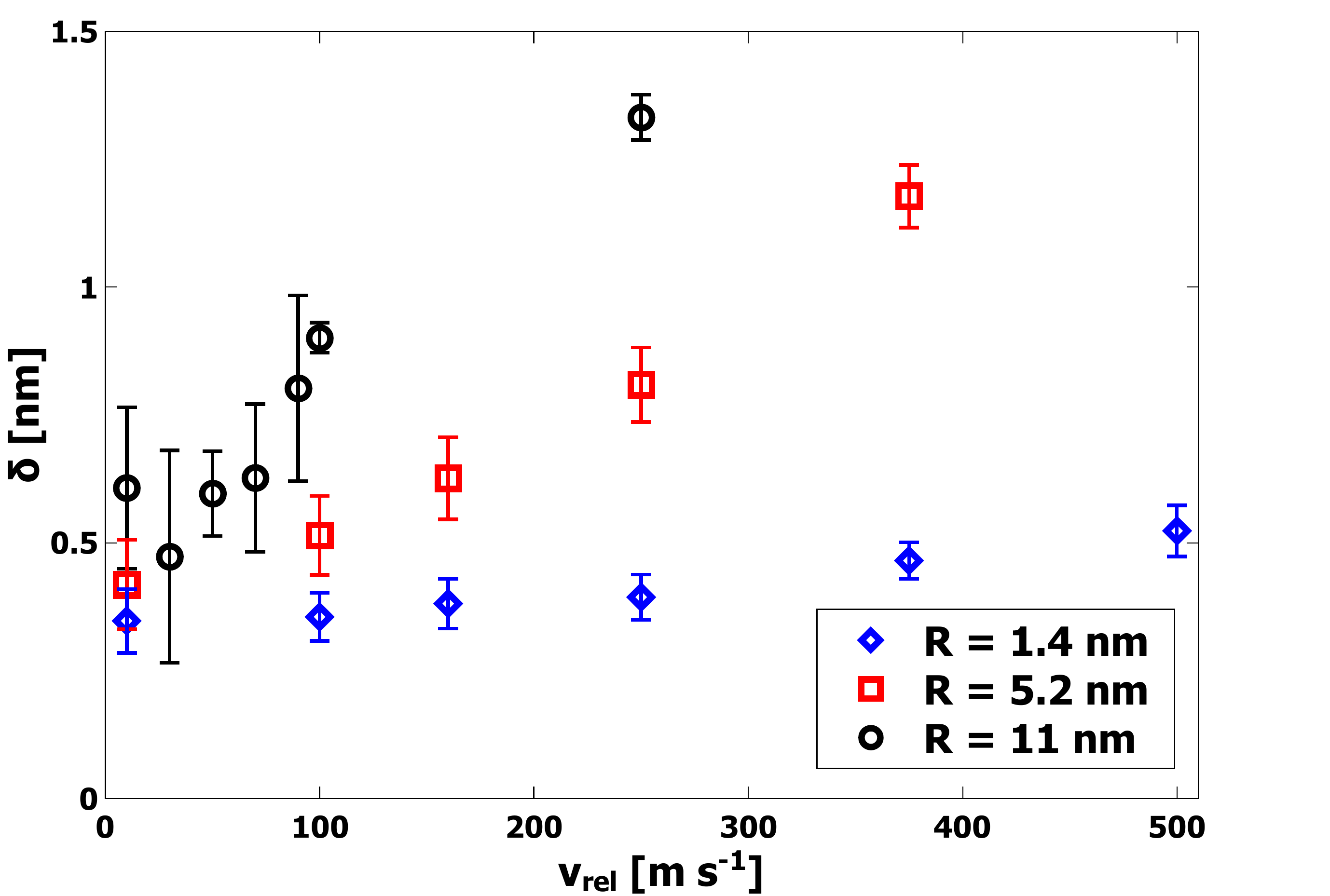}
\end{figure}

If elastic strain energy and energy associated with coordination defects are neglected, it is possible to use the computed values of $W_{adh}$ to establish an effective contact area

\begin{equation} \label{Aeff}
	A_{eff}=\frac{W_{adh}}{2 \gamma}.
\end{equation}
As in Quadery et al.\cite{quadery2017}, we also compare this effective contact area to the cross-sectional area via 

\begin{equation} \label{eta}
	\eta=\frac{A_{eff}}{\pi R^2}.
\end{equation}
The unitless parameter $\eta$ captures both the relative area of the interface as well as the quality of the bonding. Specifically, values of $\eta \approx 1$ imply a perfectly-coordinated interface with both nanoparticles deformed such that the interfacial area corresponds to the entire cross-sectional area. By contrast, values of $\eta$ that approach 0 imply a weakly bonded interface, either with a small contact area or a significant number of defects. Generally, it is expected that $\eta$ will have a value intermediate to these extremes. From the data plotted in Figure \ref{fig:etavel}, at {$v_{rel}$ = 10 m s$^{-1}$}, the value of $\eta = 0.29$ for $R = $ 1.4 nm nanoparticles indicates a substantial contact area at the interface between the nanoparticles even at the lowest values of $v_{rel}$. For larger nanoparticles, the values of $\eta$ are significantly smaller, yet still are indicative of substantial contact and strong bonding.  Additionally, the final compression length $\delta$ was directly determined from the MD simulations and Eq. \ref{deltatwor}. In Figure \ref{fig:deltavel}  $\delta$ is plotted as a function of $v_{rel}$ for each radius $R$. It is clear that $\delta$ increases strongly with increasing $v_{rel}$. This observation is consistent with plastic deformation. 

\begin{figure}
\caption{Velocities and accelerations of both nanoparticles plotted as a function of time for the R = 1.4 nm nanoparticle collision with $v_{rel}$ = 100 m s$^{-1}$. Solid red lines denote data for the first of two nanoparticles, and dashed black lines denote data for the second.  Phases and behaviors are described in the main text.}
\label{fig:velacc100}
\centering
\includegraphics[width=\linewidth]{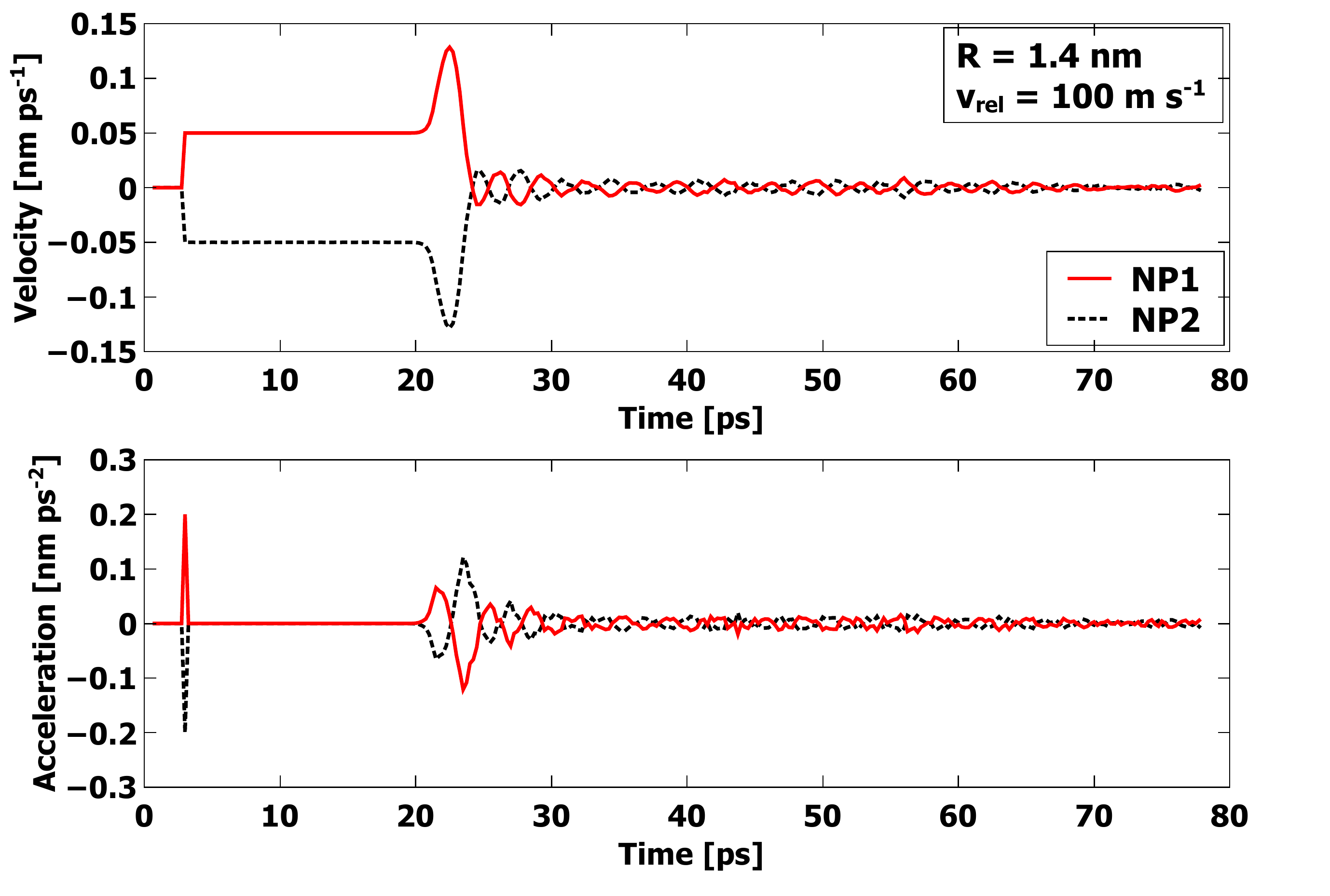}
\end{figure}

To establish the dissipation mechanism, we will now focus on the low-velocity collisions. In all collisions, the center-of-mass coordinates of each nanoparticle were retained as a function of time, allowing for determination of the acceleration and hence net force during the collision.  In Figure \ref{fig:velacc100}, the velocity and acceleration of both nanoparticles are plotted as a function of time for one simulated collision of two R = 1.4 nm nanoparticles with $v_{rel}$ = 100 m s$^{-1}$. The collision just after 20 ps is evident; the spike is an artifact of how the initial translational velocity was imparted. Due to the strong attraction, the nanoparticles initially accelerate towards each other once they enter the interaction range determined by the cutoff of the EAM potential. After significant compression, the acceleration changes sign and the velocities slow. However, before the velocities change sign for the rebound, the forces already have begun to decrease. If this were an elastic deformation, any increased compression of the two particles would lead to increased force as more elastic energy is stored. The point where the velocities change sign corresponds to the rebound phase. However, during the rebound phase, the magnitude of the acceleration is dramatically decreased from the compression phase, and consequently the rebound velocities are quite small. These observations clearly show that dissipation is correlated with strong plastic deformation, and in fact the plastic deformation itself represents the primary dissipation mechanism.

\begin{figure}
\caption{Center-of-mass kinetic energy normalized by the energy to be dissipated plotted as a function of time for all three nanoparticle sizes when $v_{rel}$ = 10 m s$^{-1}$. Blue diamonds represent data for R = 1.4 nm (solid blue line is a fit to exponential decay), red squares represent data for R = 5.2 nm (long dashed red line for fit), and black circles represent data for R = 11 nm (short dashed black line for fit). For R = 11 nm, the decay time was 27 ps. For R = 5.2 nm, the decay time was 4.6 ps. For R = 1.4 nm, the decay time was 0.7 ps.}
\label{fig:comkin}
\centering
\includegraphics[width=\linewidth]{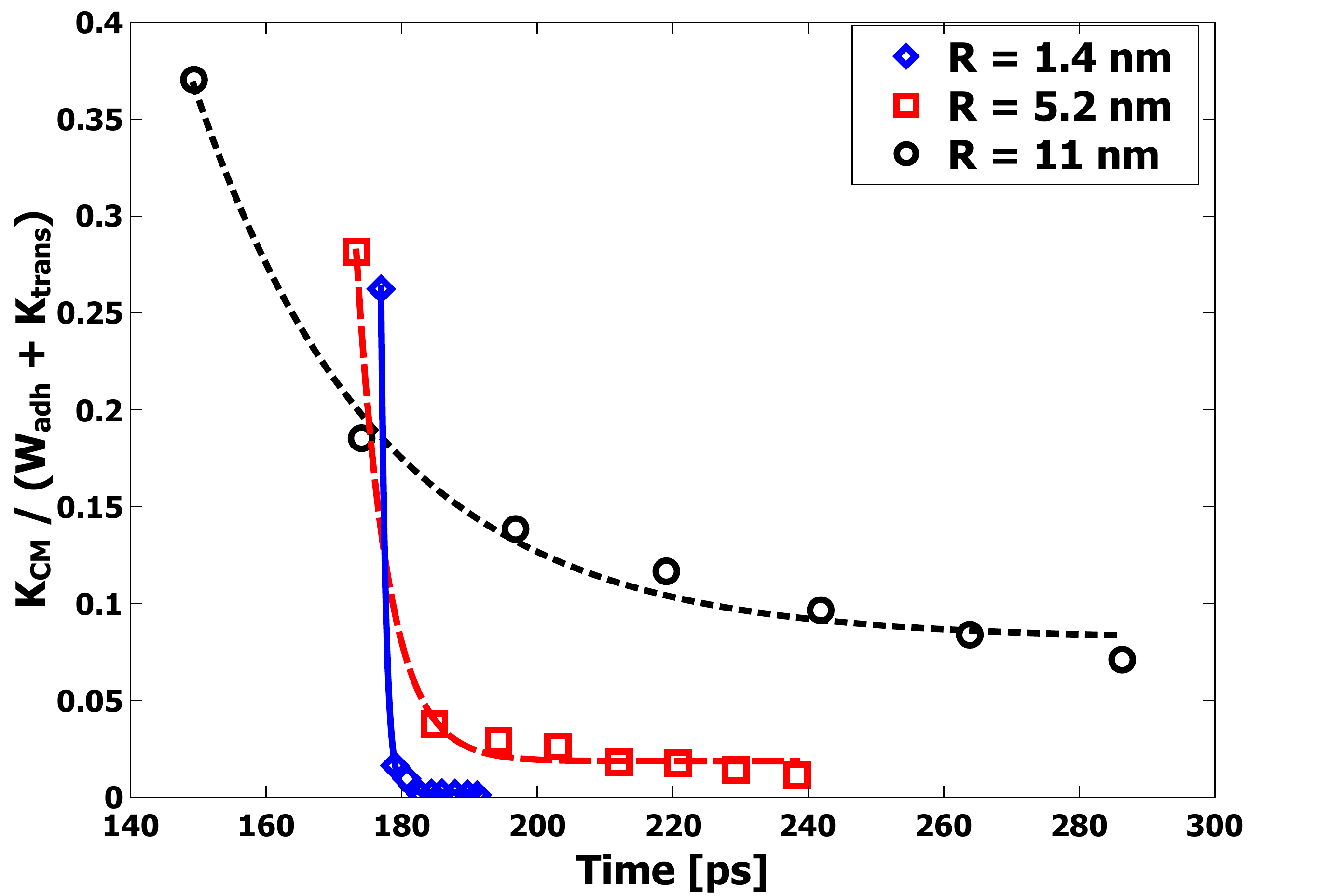}
\end{figure}

After the initial deformation and rebound phase, the subsequent behavior of the center-of-mass coordinates is consistent with that of an underdamped simple-harmonic oscillator. The results for the center-of-mass velocity (e.g. Figure \ref{fig:velacc100}) were used to establish the oscillation period $\tau$ for collisions with $v_{rel}$ = 10 m s$^{-1}$. As expected, the period increases with increasing radius $R$. In Figure \ref{fig:comkin}, the kinetic energy in the oscillations of two nanoparticles of mass $m$ and center-of-mass velocity $v_{CM}$, $K_{CM}$ $=$ $m$ $v_{CM}^2$, is normalized by the energy to be dissipated and plotted as a function of time again for $v_{rel}$ = 10 m $s^{-1}$ collisions. It is evident from Figure \ref{fig:comkin} that the energy of the vibrational motion of the particles is very strongly damped. Exponential decay function fits were performed to determine the damping time $\tau_{d}$. In Table \ref{table1}, the values of $\tau$ and $\tau_{d}$ are given for each radius $R$ for $v_{rel}$ = 10 m s$^{-1}$ collisions. For the particles to bounce, the requirement would be that $\tau_{d} >> \tau$, so that most of the incident energy is available during the rebound phase. In each case, $\tau$ is significantly greater than $\tau_d$, demonstrating very strong damping; however, it is evident that $\tau_{d}$ is increasing with $R$ faster than $\tau$, indicating the potential for bouncing at large enough $R$ values. In the next section, this condition will be explored to determine when bouncing might be expected to occur.

\begin{table}
\begin{center}
\caption{Oscillation periods $\tau$ and damping times $\tau_{d}$ for $v_{rel}$ = 10 m s$^{-1}$ as a function of nanoparticle radius $R$}
\begin{tabular}{|c|c|c|}
\hline
   $R$ (nm)& $\tau$ (ps)  & $\tau_{d}$ (ps)   \\\hline
$1.4$                     & 3.7 & 0.7\\\hline
$5.2$                     &  18.0 & 4.6  \\\hline
$11.0$                     & 45.0 & 27.0  \\\hline
\end{tabular}
\label{table1}
\end{center}
\end{table}

The stresses at the interface can be estimated from the computed forces as well as a reasonable estimate of the contact area. Using the values of $\eta$ shown in Figure \ref{fig:etavel} as an estimate of the contact area, the stresses during the collision, both tensile and compressive, are in the range of $3-10$ GPa in magnitude. This is significantly greater than the yield stress for bulk Fe of 80-100 MPa\cite{bulkfeyield}.  While yield stresses of nanoparticles can be somewhat larger than for bulk materials the differences are generally fairly small. For example, in Hawa et. al\cite{hawa2007}, simulations were used to compute the yield stresses for Ag nanoparticles, with yield stresses in the range 0.5-0.7 GPa for crystalline nanoparticles with $R \sim 5-10$ nm, and somewhat lower yield stresses for amorphous nanoparticles.  Therefore, even taking into account the higher yield stresses typically exhibited by nanoparticles, the stresses exerted upon Fe nanoparticles in the simulation are large enough to result in plastic deformation.  In Chokshi et al., the authors briefly discuss the sizes below which plastic deformation should be relevant for dissipation\cite{chokshi1993}.  Following their arguments, for the currently presented simulations plastic deformation should only be important for sizes smaller than roughly R $\sim$ 1.3 nm, which is significantly smaller than the R = 11 nm particles reported here. However, it is also clear that more accurate calculations of stress would be worthwhile, including a calculation of the stress gradients in the vicinity of the contact region.

\section{Discussion and Analysis}

The results presented above demonstrate some important features. Specifically, at least in the case of particles of $R \leq$ 11 nm, plastic deformation and extremely strong dissipation occurs in a manner not consistent with the JKR theory. Though much of the interest in particle interaction dynamics lies in larger particles of at least $R \sim 1\mu$m and beyond, where direct MD simulations are not possible, some behaviors can be predicted based on how various quantities scale with $R$. In this section we first strengthen the understanding of how JKR fails at nanometer length scales, and then explore how the results scale with radius $R$ to determine how MD predictions can be used to develop theoretical understanding at larger length scales, including where JKR likely has more validity.

In the JKR model, elastic strain accommodates an increase in the contact area. In addition, it predicts that the final contact area does not depend on the collision velocity $v_{rel}$ and that above a certain velocity, the initial translational kinetic energy is too great to cause adhesion. In order to elucidate which contributions to $U_{JKR} = U_S + U_E$ were most important in determining the value of $W_{adh}$ calculated as per Eq. \ref{wadh}, we used the directly computed values of $\delta$ to numerically solve Eqn. \ref{deltajkr} and to get an expression for contact radius $a$ as a function of $v_{rel}$.  These contact radii were then used in Eqs. \ref{jkrsurf} and \ref{jkrelastic} to obtain values for the contributions to $U_{JKR}$.  These  are plotted on Figures \ref{fig:wadhsmall}, \ref{fig:wadhmed}, and \ref{fig:wadhbig}, for comparison to the MD simulation results for $W_{adh}$. Clearly, the surface energy contribution accounts very well for the values for $W_{adh}$ for velocities up to $v_{rel}$ = 500 m s$^{-1}$.   Figure \ref{fig:deltavel} shows that $\delta$ increases with $v_{rel}$, resulting in an increase in contact radius $a$ and consequently an increased magnitude for the surface energy contribution $U_S$.  The JKR prediction for the elastic energy component $U_{E}$ is not consistent with the trend shown by $W_{adh}$. These observations present a clear picture of plastic deformation as the mechanism responsible for the increased contact area. Hence, both $\delta$ and $a$ are found to increase with increasing $v_{rel}$ in a manner which could not occur if the deformation included elastic strain energy.  In addition, the computed values of $\delta$ plotted in Figure \ref{fig:deltavel} are significantly greater than the predictions of JKR. Specifically, JKR theory predicts values of $\delta_{eq} =$ 0.23 nm, 0.35 nm, and 0.43 nm for the small, medium, and large nanoparticles, respectively, which in all cases underestimate the values shown in Figure \ref{fig:deltavel}. This demonstrates enhanced compression beyond the predictions of JKR, consistent with the observation of plastic deformation.

\begin{figure}
\caption{Computed values of $\eta$ plotted as a function of nanoparticle radius.  Red circles are for $v_{rel}$ = 10 m s$^{-1}$ and the red long dashed line shows the fitted curve.  Black circles are for $v_{rel}$ = 100 m s$^{-1}$ and the black short dashed line shows the fitted curve.  Blue diamonds are for $v_{rel}$ = 250 m s$^{-1}$ and the blue solid line shows the fitted curve.  }
\label{fig:etasize}
\centering
\includegraphics[width=\linewidth]{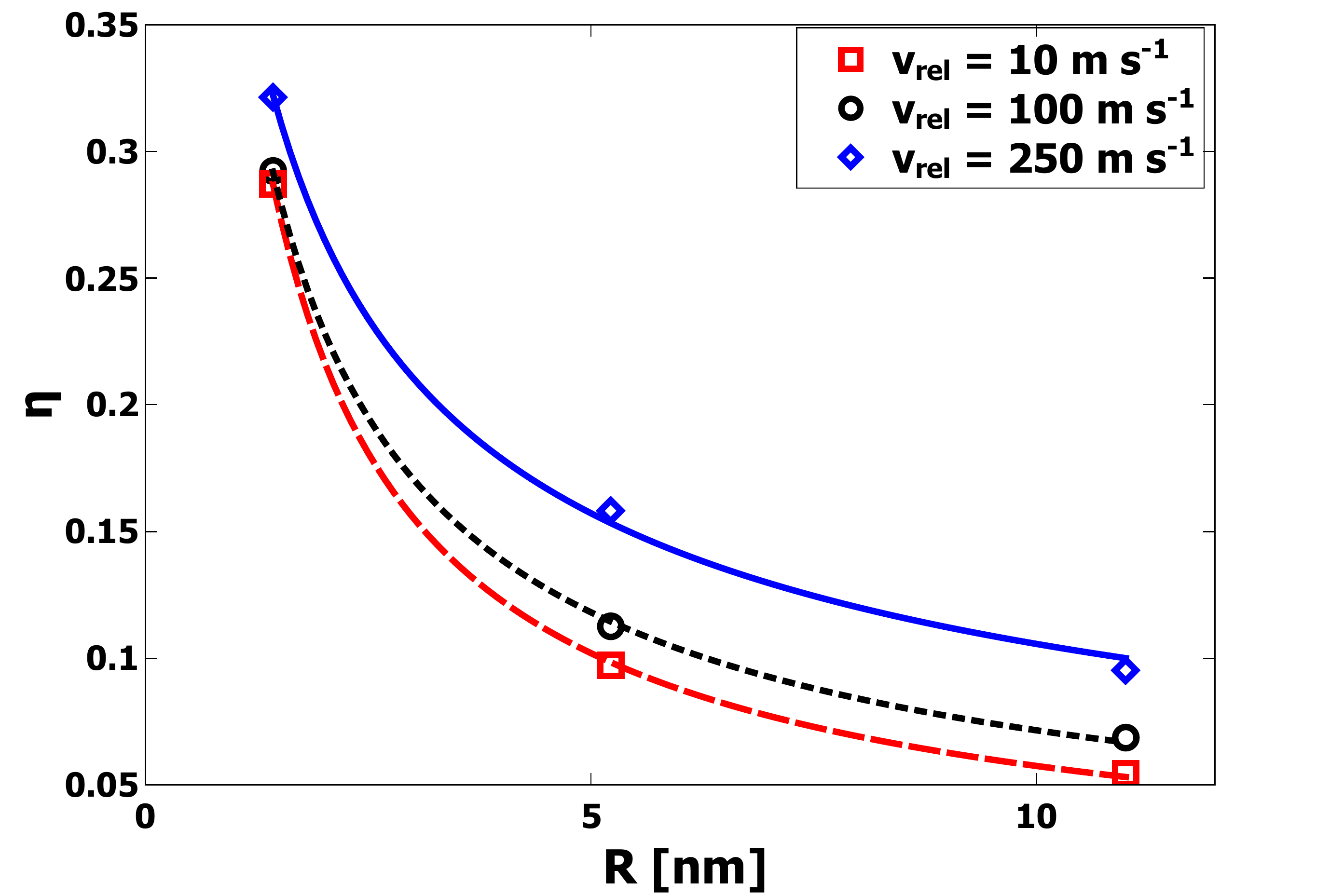}
\end{figure}

In Figure \ref{fig:etasize}, we plot values of $\eta$ for three velocities as a function of nanoparticle radius $R$. The gradual increase in $\eta$ with $v_{rel}$ is consistent with the observation of greater plastic deformation. The decrease in $\eta$ with increasing radius can be understood to be in part simply a geometric effect. Because the interactions are finite range, when $R$ increases significantly beyond 0.53 nm, the cut off distance for interactions, simple geometric arguments suggest that $\eta$ should scale as $R^{-1}$. The actual data shows a trend with a somewhat different scaling exponent, likely due to the fact that the strongest plastic deformation happens for the smallest nanoparticles.  The value of $\eta$ appears to scale with radius $R$ approximately as $R^{-0.83}$ for $v_{rel} =$ 10 m s$^{-1}$. For higher $v_{rel}$, the scaling changes somewhat. Specifically, for $v_{rel} = 100$ m s$^{-1}$ , $\eta$ scales as $R^{-0.72}$, while for $v_{rel}=250$ms$^{-1}$ , $\eta$ scales as $R^{-0.57}$.  The dependence on $v_{rel}$ is clearly due to the fact that $K_{trans}$ becomes dominant in comparison to surface interaction as $v_{rel}$ increases. For low enough velocities the contact area indicated by $\eta$ appears to increase approximately linearly with $R$. The relevance for collisions at very large scale, both in terms of the dissipation mechanism and the crossover towards bouncing behavior, will be addressed in the final section. However, we note that for $v_{rel}$ = 10 m $s^{-1}$ and  R = 1 $\mu m$, the scaling behavior results in a prediction $\eta \approx 1.2 \times 10^{-3}$, which indicates that the adhesion and dissipation occurs over a much smaller relative area than for $R = 11$ nm and smaller particles.

\begin{figure}
\caption{Final collision velocity $v_{rel,f}$ plotted as a function of initial kick velocity $v_{rel}$ for all three nanoparticle sizes.  This figure clearly demonstrates the existence of a minimum collision velocity due to surface attraction effects.  Values are denoted by blue diamonds for R = 1.4 nm, red squares for R = 5.2 nm, and black circles for R = 11 nm; the fitted curves are denoted by a blue short dashed line for R = 1.4 nm, a red long dashed line for R = 5.2 nm, and a black solid line for R = 11 nm.}
\label{fig:vrelf}
\centering
\includegraphics[width=\linewidth]{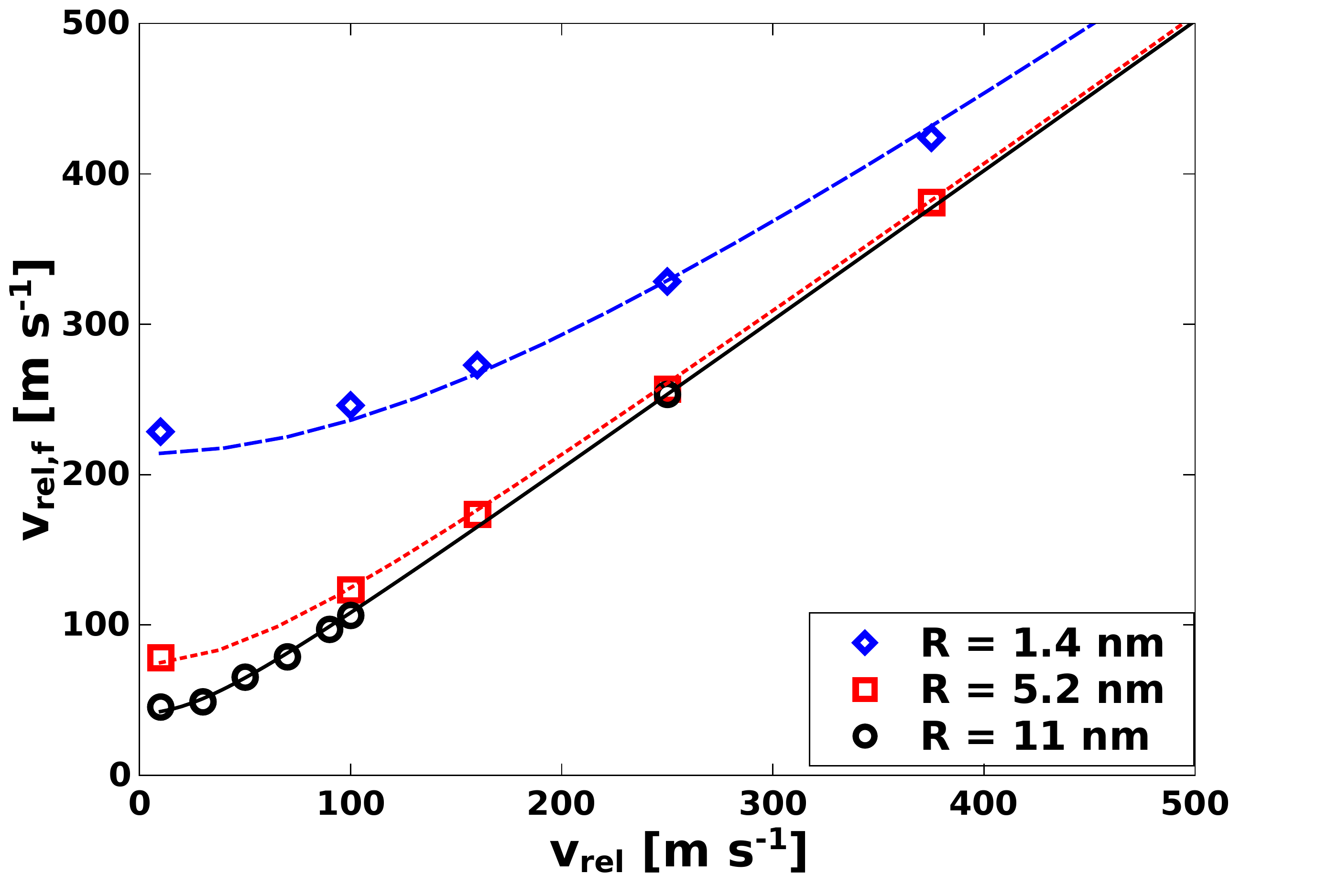}
\end{figure}

For small particles $R = 11$ nm and below, the results indicate that plastic deformation always occurs. Indeed, we observe that is not possible at these scales to lower $v_{rel}$ sufficiently to observe elastic behavior consistent with JKR theory. For small particles, surface attraction energy dominates $K_{trans}$ for the lower values of $v_{rel}$. The result of the strong attraction is that the nanoparticles accelerate significantly just before the collision, and the effective collision velocity $v_{rel,f}$ can often be substantially greater than the initial velocity $v_{rel}$.  Assuming that the surface interaction is conservative, then the effective collision velocity $v_{rel,f}$ should depend on the initial velocity $v_{rel}$ and the size-dependent velocity $v_c$,

\begin{equation} \label{vc}
	v_{rel,f}=v_c \Big[ 1 + \Big( \frac{v_{rel}}{v_c} \Big)^2 \Big]^{1/2}.
\end{equation}
The value of $v_{c}$ is a parameter which depends on $R$ and was determined by examination of the maximum nanoparticle velocity during the collision.
In Figure \ref{fig:vrelf},  $v_{rel,f}$ obtained from simulation is plotted as a function of $v_{rel}$ along with the fit curve from Eq. \ref{vc}.  All data was used to obtain fits, but only a subset is plotted for clarity. For the three radii $R =$ 1.4 nm, $R =$ 5.2 nm, and $R =$ 11 nm, the values for the fitted parameter $v_c$ are respectively $v_c = $ 213.9 $\pm$ 4.9 m s$^{-1}$, $v_c = $ 74.3 $\pm$ 1.7 m s$^{-1}$, and $v_c = $ 41.2 $\pm$ 1.5 m s$^{-1}$. This demonstrates that, for these small particle sizes, the velocity at collision $v_{rel,f}$ is substantially greater than the lowest value of $v_{rel}$ simulated. Consequently, a simulation with a very small $v_{rel}$ will result in $v_{rel,f} \approx v_{c}$ as a minimum effective collision velocity.  Therefore, while it might be thought that a low enough value of $v_{rel}$ should exist where collisions are elastic, the present results demonstrate that for $R=11$nm and below, this is not the case. Specifically,  for $R=11$nm,  $v_c = $ 41.2 $\pm$ 1.5 m s$^{-1}$ is substantially greater than the lowest $v_{rel}$ = 10 m s$^{-1}$, and even much lower values of $v_{rel}$ would yield essentially the same collisions with plastic deformation. For smaller particles surface attraction is even more important, and the very large values of $v_{c}$ result in even more dramatic plastic deformation at all values of $v_{rel}$.
In understanding scaling behavior, we note that $v_{c} \propto R^{-0.81}$. This indicates that as $R$ increases, surface attraction becomes a less significant factor. However, even for $R$ = 1 $\mu$m particles, the scaling of $v_{c}$ predicts $v_{c} \approx$ 1 m s$^{-1}$. Therefore, even if $v_{rel}$ is below 1 m s$^{-1}$, the effective collision velocity will be $v_{rel,f} \approx$ 1 m s$^{-1}$, which is still substantial and should involve some plastic deformation at the interface. This is further discussed in the last section of the paper.

It is possible to use scaling relations to predict the size $R$ where bouncing will occur. Specifically, one criterion for bouncing is that the kinetic energy during the first rebound should be larger than $W_{adh}$ in order for the particles to separate. We apply this criterion for $v_{rel}$ = 10 m s$^{-1}$ collisions. The results above for $\eta$ demonstrate that $W_{adh} \propto R^{1.1}$, whereas the kinetic energy during the first rebound scales $K \propto R^{3.4}$, hence eventually the rebound kinetic energy will be substantially greater than $W_{adh}$. For $v_{rel}$ = 10 m s$^{-1}$, this criterion results in the prediction that bouncing might occur for $R \approx$ 23 nm. However, the fact that substantial plastic deformation occurs casts some doubt on these results, since the neck formed by the deformed surfaces means the behavior is strongly irreversible. In fact, as the results plotted in Figures \ref{fig:wadhsmall}, \ref{fig:wadhmed}, and \ref{fig:wadhbig}, demonstrate, even negative values of $W_{adh}$ occur without bouncing, due to the very strong deformation of the particles.

\begin{figure}
\caption{Plot of $\tau$ and $\tau_{d}$ as a function of nanoparticle radius $R$ for $v_{rel} =$ 10 m s$^{-1}$. }
\label{fig:tau}
\centering
\includegraphics[width=\linewidth]{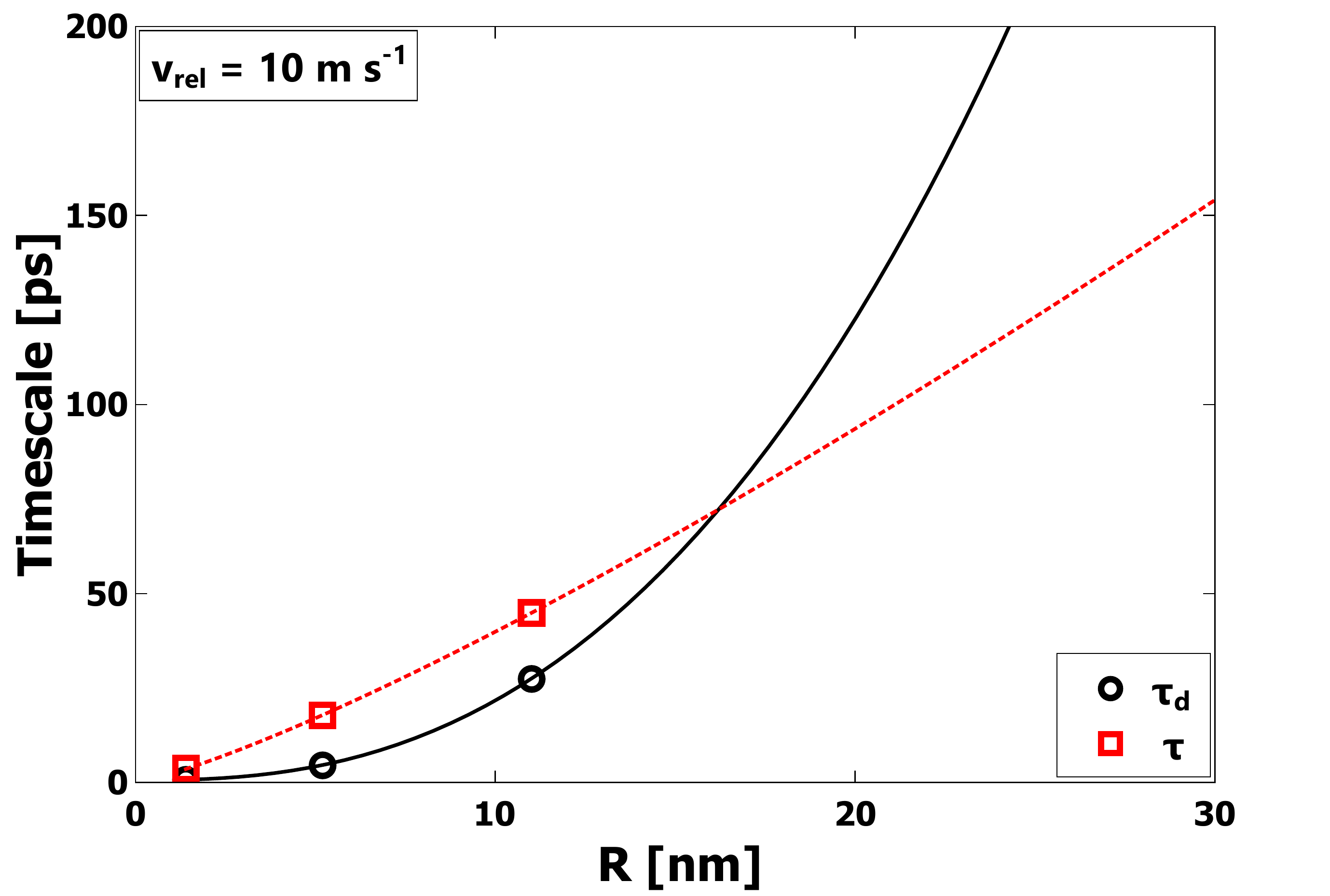}
\end{figure}

Another approach to predict behavior at larger scales is to use the computed scaling of the oscillation period $\tau$ and the damping time $\tau_{d}$. As described above, bouncing would seem to require $\tau_{d} >> \tau$ in order for bouncing to occur, since strong dissipation of the kinetic energy of the two particle oscillations prevents separation of the particles. In Figure \ref{fig:tau}, the scaling of $\tau$ and $\tau_{d}$ with particle size $R$ is shown. From Figure \ref{fig:tau}, the fit indicates $\tau \propto R^{1.23}$, and $\tau_{d} \propto R^{2.50}$, which leads to a crossover at $R=17.3$nm. Beyond $R \approx$ 20 nm, $\tau_{d}$ eventually becomes substantially greater than $\tau$ and eventually bouncing should occur.

\section{Conclusions}

For metallic nanoparticles of radii $R =$ 11 nm and below, simulation results indicate adhesion occurs in every head-on collision. Simulation results show extremely large stress values and strong plastic deformation, indicating that JKR is not applicable in the present case. Because the observed deformation is plastic, it does not include large elastic strain energy contributions predicted by JKR, thereby resulting in a larger contact area that increases strongly with increasing $v_{rel}$. Moreover, the primary mechanism for the dissipation appears to be connected strongly to the atomic rearrangement that occurs at the interface associated with the plastic deformation. Scaling behavior suggest that for relatively low collision velocities, $v_{rel} \sim$ 10 m s$^{-1}$, nanoparticles would need to be at least $R \sim$ 20 nm in order for bouncing to occur.

For nanoparticles, surface attraction can often dominate incident kinetic energy, and plastic deformation is expected to always occur for any $v_{rel}$. The results show that even for $R \approx 1\mu m$, the relative velocity during the collision is at least $v_{rel,f} \approx$ 1 m s$^{-1}$. Using the scaling of $v_{c} \propto R^{-0.81}$ at $v_{rel} =$ 10 m s$^{-1}$, and the fact that the mass of a grain scales as $R^{3}$, the minimum kinetic energy in a collision scales as $R^{1.38}$. 
However, the scaling of $\eta$ at the same $v_{rel}$ shows that the kinetic energy associated with $v_{c}$ needs to be dissipated over an effective area which scales as $R^{1.1}$. This suggests that the minimum kinetic energy associated with $v_{c}$ increases more rapidly with $R$ than the area available to dissipate the energy. Therefore, it is quite possible that while dissipation becomes less effective as $R$ increases, and as expected bouncing becomes the dominant behavior, strong plastic deformation at the contact interface likely occurs. Hence, while $v_{c}$ decreases with $R$, the kinetic energy associated with $v_{c}$ actually increases, and because the relative area for the collision tends to decrease, it is reasonable to expect high stress and plastic deformation at the contact. However, this possibility  remains a subject requiring more direct verification.

There is a direct relationship to the conventional explanation of Amonton's laws of friction, wherein dissipative frictional forces are independent of apparent contact area and are solely dependent on the normal force at an interface. This has been contradicted based on the presence of nanoscale asperities\cite{bhushan2017} which result in an actual contact area which is generally much smaller than the apparent contact area. When normal forces exist, either due to surface attraction or some other applied force, the actual contact area grows often due to plastic deformation of the asperities, and hence the frictional force increases. When the actual contact area is much smaller than the apparent contact area, the stresses in the asperities can become quite large. This is very similar to the results found here. Specifically, since $\eta \propto R^{1.1}$ is less than $R^{2}$, the predictions here indicate that as $R$ increases, the stress at the contact point will tend to increase quite dramatically, thereby leading to plastic deformation. However, the results also show that any enhanced dissipation with increasing $R$ is not enough to prevent bouncing behavior at larger values of $R$.

In addition, the strong attraction which results in the larger effective collision velocities is the same reason plastic deformation occurs. In other words, even when the incident kinetic energy is relatively low, strong interaction tends to result in plastic deformation.  As $R$ increases, surface attractive forces do not become weaker. Instead, the area where strong interactions occur increases less rapidly with $R$ than the overall mass of the particles does, and hence the value of $v_{c}$ decreases with increasing $R$. However, because the interactions are localized over a smaller relative area, they can still be very strong even when $v_{c}$ is small. In fact, we expect that not only stresses at the contact are very strong, but that very strong stress gradients likely are responsible for the observed plastic deformation. This point will be a focus of future efforts.

While the general picture of plastic deformation as the dominant mechanism for adhesion and dissipation is in stark contrast to JKR, some features remain consistent. Specifically, the simulations demonstrate the formation of a ``neck'' which tends to increase the adhesion and more strongly prevent rebound. The distinction is in the mechanism for the formation of the neck, which we find to be plastic deformation rather than elastic deformation. This view is also consistent with previous efforts in simulations of silica particles which demonstrated strong plastic deformation \cite{urbassek2017,quadery2017}. It should also be noted that other works have explored viscoelastic dissipation mechanisms associated with plastic deformation\cite{krijt2013}, even though it appears that the dependence of the contact area on $v_{rel}$ and $R$ has not been previously considered.  These insights could be of particular interest to the nuclear and pharmaceutical industries, where critical processes depend on tightly controlled powder flows. Additionally, the adhesion of nanoparticles could have a significant effect on their catalytic performance via a reduction in surface area.

Future work will be directed towards elucidating behavior of silicate and other oxide particles, using some of the same approaches here. It is expected that surface bonding occurs with more defects when cations and anions are present, and also that plastic deformation by the motion and generation of dislocations at the interface occurs less readily than with metals. It is also possible that the approach of determining the oscillation and damping times $\tau$ and $\tau_{d}$ can be extended to computation of the coefficient of restitution when bouncing occurs.

\section*{Acknowledgments}

This research was made possible by support from the National Science Foundation, Division of Astronomical Sciences,  under grant 1616511. The authors also acknowledge support and computational time provided by the Institute of Simulation and Training and the STOKES computer cluster at the University of Central Florida.

\section*{Data availability}
 The raw and processed data required to reproduce these findings cannot be shared at this time as the data also forms part of an ongoing study.

\section*{Bibliography}
\bibliography{sources.bib} 
\bibliographystyle{unsrt}
\end{document}